\input amstex
\magnification 1200
\TagsOnRight
\def\qed{\ifhmode\unskip\nobreak\fi\ifmmode\ifinner\else
 \hskip5pt\fi\fi\hbox{\hskip5pt\vrule width4pt
 height6pt depth1.5pt\hskip1pt}}
\NoBlackBoxes
\hyphenation{Mc-Graw}
\baselineskip 18 pt
\parskip 6 pt

\centerline {\bf INVERSE SCATTERING ON THE LINE FOR}
\centerline {\bf A GENERALIZED NONLINEAR SCHR\"ODINGER EQUATION}

\vskip 10 pt
\centerline {Tuncay Aktosun}
\vskip -8 pt
\centerline {Department of Mathematics and Statistics}
\vskip -8 pt
\centerline {Mississippi State University}
\vskip -8 pt
\centerline {Mississippi State, MS 39762, USA}
\vskip -8 pt
\centerline {aktosun\@math.msstate.edu}

\centerline {Vassilis G. Papanicolaou and Vassilis Zisis}
\vskip -8 pt
\centerline {Department of Mathematics}
\vskip -8 pt
\centerline {National Technical University of Athens}
\vskip -8 pt
\centerline {Zografou Campus}
\vskip -8 pt
\centerline {157 80, Athens, Greece}
\vskip -8pt
\centerline {papanico\@math.ntua.gr}

\noindent {\bf Abstract}: A one-dimensional generalized nonlinear Schr\"{o}dinger equation
is considered, and the corresponding inverse scattering problem is
analyzed when the potential is compactly supported
and depends on the wave function. The unique recovery of the potential is
established from an appropriate set of scattering data.

\vskip 15 pt
\par \noindent {\bf PACS (2003):} 2.30.Zz, 3.65.Nk, 43.25.+y 
\vskip -8 pt
\par \noindent {\bf Mathematics Subject Classification (2000):}
34A55, 34L25, 34L30
\vskip -8 pt
\par\noindent {\bf Keywords:} nonlinear potential, generalized nonlinear
Schr\"{o}dinger equation, inverse scattering, nonlinear scatterer

\newpage

\noindent {\bf 1. INTRODUCTION}
\vskip 3 pt

Consider the nonlinear equation
$$-u''+Q(x,u)\,u=k^{2}u,\qquad x\in {\bold R},\tag 1.1$$
where $k$ is a real parameter, the prime denotes the derivative
with respect to the spatial variable $x$, and $Q(x,u)$ has the
form
$$Q(x,u)=\sum_{n=0}^{\infty }q_{n}(x)\,u^{n},\tag{1.2}$$
with each $q_{n}(x)$ being real valued, bounded, measurable,
supported in $[0,b]$ for a fixed $b>0,$ and the series
$$\sum_{n=0}^{\infty }\left(\sup_{x\in [0,b]}
|q_{n}(x)| \right)u^{n}\tag{1.3}$$
being entire in $u$. The assumption that the series in (1.3) is entire
is equivalent to assuming that its radius of convergence is infinite, i.e.
$$\varlimsup_{n\to+\infty}\left(\sup_{x\in [0,b]}|q_{n}(x)|
\right)^{1/n}=0.\tag{1.4}$$
Note that such an assumption is stronger than just assuming that $Q(x,u)$
given in (1.2) is entire in $u$.

In this paper we consider the solution to (1.1) satisfying
$$u(0;k)=\varepsilon,\qquad u^{\prime}(0;k)=-ik\varepsilon.\tag{1.5}$$
As shown in Proposition~2.1, a unique solution to (1.1) exists for
all $x\in{\bold R}$ and $k\in{\bold R}$ when $|\varepsilon|$ is
sufficiently small. The results given in our
paper hold both for real and complex values of
$\varepsilon.$ We suppress the dependence of $u$ on
$\varepsilon$ for simplicity. Since for $n\geq 0$ we assume
$q_{n}(x)=0$ when $x\notin [0,b],$ it follows that for each
$k\in{\bold R}\setminus \{0\}$ the general solution to (1.1) for
$x\notin (0,b)$ is a linear combination of $e^{ikx}$ and
$e^{-ikx}$. Thus, we have
$$u(x;k)=\varepsilon e^{-ikx},\qquad x\leq 0,\tag{1.6}$$
which is equivalent to (1.5), and
$$u(x;k)=A(k;\varepsilon)\,e^{ikx}+B(k;\varepsilon)\,
e^{-ikx},\qquad x\geq b.\tag{1.7}$$

Because of the resemblance with the (linear) Schr\"{o}dinger equation, we
will refer to (1.1) as a generalized nonlinear Schr\"{o}dinger equation
and to $Q(x,u) $ as the (nonlinear) potential.
The expressions for $u(x;k)$ given in (1.6) and (1.7) indicate that
we have a scattering problem in hand, where a plane wave is sent from $%
x=+\infty $ onto the nonhomogeneity $Q(x,u)$, and a part of the wave is
transmitted to $x=-\infty $ and a part is reflected back to $x=+\infty $.
However, contrary to the linear case, there is in general
no energy conservation, i.e. $|B(k;\varepsilon)|^2-|A(k;\varepsilon)|^2$
in general depends on $k.$
The linear part $q_0(x)$ of the potential $Q(x,u)$ in (1.2)
represents the restoring force density in wave propagation
governed by (1.1) in the frequency domain. The higher order
terms in the potential allow the description of nonlinear reaction
of the  medium to propagation of
elastic waves.

In analogy with
the direct and inverse scattering problems for the linear Schr\"{o}dinger
equation, the corresponding problems for (1.1) can be formulated as
follows. In the direct problem, given $Q(x,u)$, our task is to determine the
``scattering coefficients'' $A(k;\varepsilon)$ and $B(k;\varepsilon)$ for
sufficiently small $|\varepsilon|.$ On the other hand, in the inverse
scattering problem, given some scattering data related to $A(k;\varepsilon) $
and $B(k;\varepsilon)$, the task is to recover $Q(x,u)$. In our paper, we do
not study the characterization of the scattering data
so that the existence of a corresponding potential is assured;
we only discuss the uniqueness and
recovery aspects of our inverse problem by assuming that there exists
at least one potential corresponding to our scattering coefficients.

The inverse scattering problem
analyzed in this paper is analogous to the recent study by Weder [1],
where the time-dependent Schr\"{o}dinger equation with a nonlinear term is
investigated and both the linear and nonlinear parts of the potential are
recovered from some appropriate scattering data by using a time-domain method.
For  other related studies of inverse problems on nonlinear equations using time-domain
methods, we refer the reader to [2-5] and the references therein.

Our paper is organized as follows. In Section~2 we show that the direct
and inverse scattering problems
for the nonlinear equation (1.1) are equivalent to the corresponding
problems for an infinite
number of linear equations, and we analyze
the basic properties of the scattering data for each linear equation
and define the appropriate data set $\Cal D_n$ given in (2.16) for each $n\ge 1.$
We then solve the inverse scattering problem for each $n\ge 1$
recursively. The solution of the inverse problem when
$n=1$ is well known, and in Section~3 we list
the basic facts from the case $n=1$
that are needed later on to solve the inverse problems for
$n\ge 2.$ In Section~4 we show that the solution of
the inverse scattering problem for each
$n\ge 2$ can be obtained by inverting either of the two
integral equations (4.4) and (4.5).
In Section~5 we prove the unique invertibility of (4.4) and (4.5) to recover
$q_{n-1}(x)$ for each $n\ge 2,$ and we summarize the recovery
of the nonlinear potential $Q(x,u)$ in terms of the scattering
data involving $A(k;\varepsilon)$ and $B(k;\varepsilon).$
Finally, in Section~6, we illustrate the direct and inverse problems
for (1.1) with some concrete examples.

\vskip 10 pt
\noindent {\bf 2. PRELIMINARIES}
\vskip 3 pt

It is straightforward to verify that $u(x;k)$ satisfies (1.1) and (1.5)
if and only if it satisfies the integral equation
$$u(x;k)=\varepsilon\, e^{-ikx}+\frac{1}{k}\int_{0}^{x}
\sin\left(k(x-t)\right)\,Q\left(t,u(t;k)\right)\,u(t;k)\,dt.\tag{2.1}$$

\noindent {\bf Proposition 2.1.} {\it There exists a constant $\delta>0$ depending only on
$Q(x,u)$, but not on $k$ (as long as $k$ is real), such that if $|\varepsilon
|\leq \delta $ then a solution $u(x;k)$ to (1.1) satisfying (1.5)
exists for all $x\in {\bold R}$, and it is unique.}

\noindent PROOF: Suppose that a solution $u(x;k)$ to (2.1) ceases to exist,
i.e. blows up, for some $x\in (0,b]$. Then, for any $r$ sufficiently large,
there is an $x_{0}\in (0,b]$ such that $|u(x_{0};k)|=r$ and $|u(x;k)|<r$ for
$x<x_{0}$. Fix one such $r$. Our assumption in (1.4) implies that there
is a $C>0$ such that $|Q(x,u)|\leq C$
for $x\in[0,b]$ and $|u|\leq r$. Let us take
$$|\varepsilon |\leq \delta :=(r/2)\,e^{-Cb^{2}}.\tag{2.2}$$
Using $|\sin \theta |\leq |\theta |$ for real $\theta $ and the
realness of $k,$ for $x<x_{0}$ from (2.1) we get
$$\aligned
|u(x;k)| &\leq |\varepsilon |+\int_{0}^{x}( x-t)\, \left|Q(t,u( t;k))
\right|\,|u( t;k)| \,dt \\
&\leq |\varepsilon |+b\int_{0}^{x}\left|Q(t,u(t;k))\right|\,| u(t;k)|\,dt.
\endaligned$$
Applying Gronwall's inequality (see, e.g. Prob. 1 in Ch. 1 of [6]) on
the last term above, we obtain
$$|u(x;k)| \leq |\varepsilon |+b\,|\varepsilon |\int_{0}^{x}
\left| Q(t,u(t;k))\right|\, \exp \left( b\int_{t}^{x}Q(z,u(z;k))\,dz
\right)\,dt.\tag{2.3}$$
Setting $x=x_{0}$ in (2.3), we have
$$r\leq |\varepsilon |+b\,|\varepsilon |\int_{0}^{x_{0}}C\exp \left(
Cb\,(x_{0}-t)\right) \,dt\leq |\varepsilon |\,e^{Cb^{2}},$$
and a comparison with (2.2) indicates that
$$0<r\leq |\varepsilon |\,e^{Cb^{2}}\leq (r/2)\,e^{-Cb^{2}}e^{Cb^{2}}=r/2,$$
which is impossible. Therefore, $u(x;k)$ does not blow up in $[0,b]$ and
hence it exists for all $x\in {\bold R}$.
The uniqueness of $u(x;k)$ follows from the Lipschitz property of $Q(x,u)$
with respect to $u$, which, in turn, follows from the analyticity
of $Q(x,\cdot )$. \qed

 From the conditions in (1.5), with the help of Theorem~8.4 in Sec. 1.8 of
[6], when $k$ is real and bounded, $x\in {\bold R}$, and $|\varepsilon
|$ is sufficiently small, we see that $u$ is analytic in $\varepsilon $ and
hence
$$u(x;k)=\sum_{n=1}^{\infty }\varepsilon ^{n}u_{n}(x;k)=\varepsilon\,
u_{1}(x;k)+\varepsilon ^{2}\,u_{2}(x;k)+\dots.\tag{2.4}$$
Note that the $\varepsilon^{0}$-term is absent in (2.4) because (1.5)
implies that $u(x;k)\equiv 0$ if $\varepsilon =0$.

We observe from (1.7) and (2.4) that the analyticity of $u$ in $\varepsilon $
for $x\geq b$ implies that $A(k;\varepsilon) $ and $B(k;\varepsilon) $
are analytic in $\varepsilon$ at $\varepsilon =0$ for
real, nonzero, and bounded $k$. Thus, we have the expansions
$$A(k;\varepsilon) =\sum_{n=1}^{\infty }\varepsilon ^{n}A_{n}(k),\quad
B(k;\varepsilon) =\sum_{n=1}^{\infty }\varepsilon^{n}B_{n}(k),\tag{2.5}$$
where we emphasize that the $\varepsilon^{0}$-terms are absent. With the
help of (2.1) and the expansions (2.4) and (2.5), one can show
that the generalized nonlinear Schr\"{o}dinger equation (1.1), the
condition (1.6), and the expression (1.7) are equivalent to an
infinite number of scattering problems for linear differential equations.
The use of (2.1) allows
us to avoid the interchange of the $x$-differentiation and the infinite
summation in (2.4). The resulting scattering problems for linear
equations are, for $n=1$,
$$-u_{1}^{\prime \prime }+q_{0}(x)\,u_{1}=k^{2}u_{1},
\qquad x\in {\bold R},\tag{2.6}$$
$$u_{1}(x;k)=e^{-ikx},\qquad x\leq 0,\tag{2.7}$$
$$u_{1}(x;k)=A_{1}(k)\, e^{ikx}+B_{1}(k)\,e^{-ikx},\qquad x\geq b,\tag{2.8}$$
and, for $n\geq 2$,
$$-u^{\prime\prime}_{n}+q_{0}(x)\,u_{n}=k^{2}u_{n}-g_{n}(x;k),\qquad x\in
{\bold R},\tag{2.9}$$
$$u_{n}(x;k)=0,\qquad x\leq 0,\tag{2.10}$$
$$u_{n}(x;k)=A_{n}(k)\, e^{ikx}+B_{n}(k)\, e^{-ikx},\qquad x\geq b,\tag{2.11}$$
where we have defined
$$g_{n}(x;k):=q_{n-1}(x)\,u_{1}(x;k)^{n}+h_{n}(x;k),\qquad n\ge 2,\tag{2.12}$$
$$h_2(x;k):=0;\quad h_{n}(x;k):=\sum_{j=2}^{n-1}C_{jn}(x;k)
\,q_{j-1}(x),\qquad n\ge 3,\tag{2.13}$$
with $C_{jn}(x;k)$ for $2\leq j\leq n-1$ being the coefficient of
$\varepsilon^{n}$ in the expansion of
$$\left[\varepsilon\, u_{1}(x;k)+\varepsilon^{2}u_{2}(x;k)+\dots
+\varepsilon^{n-1}u_{n-1}(x;k)\right]^{j}.$$
We list the first few $h_{n}(x;k)$ below:
$$h_{2}(x;k)=0,\quad h_{3}(x;k)=2u_{1}u_{2}q_{1},\tag 2.14$$
$$h_{4}(x;k)=(2u_{1}u_{3}+u_{2}^{2})q_{1}+3u_{1}^{2}u_{2}q_{2},$$
$$h_{5}(x;k)=2(u_{1}u_{4}+u_{2}u_{3})q_{1}+
3(u_{1}u_{2}^{2}+u_{1}^{2}u_{3})q_{2}+4u_{1}^{3}u_{2}q_{3}.$$

So far we have established the validity of $u_{n}(x;k)$ only for real and
bounded $k$ [cf. (2.4)] and those of $A_{n}(k)$ and $B_{n}(k)$ [cf. (2.5)]
only for real, nonzero and bounded $k $. In the next result,
we analyze their analytic extensions in $k$ to the entire complex
plane ${\bold C}$.

\noindent {\bf Proposition 2.2.} {\it For $n\geq 1,$ let
$u_{n}(x;k)$, $A_{n}(k)$, and $B_{n}(k)$
be the quantities given in (2.4) and (2.5). Then,
$u_{n}(x;k)$ for each $x\in {\bold R},$ $kA_{n}(k),$
and $kB_{n}(k)$ are entire in $k$.}

\noindent PROOF: Since $u_{1}(x;k)$ satisfies the linear equation
(2.6) with the condition in (2.7) and $u_{n}(x;k)$
for $n\geq 2$ satisfies the linear equation (2.9)
with the condition in (2.10), it follows from Theorem~8.4 in Sec. 1.8 of [6]
that $u_{n}(x;k)$ and $u^{\prime}_{n}(x;k)$ are entire in $k$. From (2.8)
and (2.11), for $x\geq b$ and $n\geq 1$ we get
$$u_{n}(x;k)=A_{n}(k)\, e^{ikx}+B_{n}(k)\, e^{-ikx},\quad u^{\prime}_{n}
(x;k)=ik\,A_{n}(k) e^{ikx}-ik\,B_{n}(k) e^{-ikx},$$
and hence
$$k\,A_{n}(k)=\frac{k\,u_{n}(x;k)-i\,u^{\prime}_{n}(x;k)}{2}\,e^{-ikx},\quad
k\,B_{n}(k)=\frac{k\,u_{n}(x;k)+i\,u^{\prime}_{n} (x;k)}{2}\,e^{ikx},\tag{2.15}$$
 from which we see that $kA_{n}(k)$ and $kB_{n}(k)$ are entire in $k$. \qed

\noindent {\bf Proposition 2.3.} {\it For each fixed $k\in {\bold C}$ and all $n\geq 1$, the
quantities $u_{n}(\cdot;k)$ are bounded for $x\in [0,b]$. For each fixed $k\in
{\bold C}$ and all $n\geq 2$, the quantities $h_{n}(\cdot;k)$ defined
in (2.13) are bounded in $x$ and vanish for $x\notin [0,b]$.}

\noindent PROOF: The boundedness of $u_{n}(\cdot ;k)$ for $x\in [0,b]$ can be
established recursively for $n\geq 1$ by using the fact that each $%
u_{n}(\cdot ;k)$ is a solution to a linear, ordinary differential equation
with appropriate initial conditions at $x=0$ [cf. (2.7) and (2.10)].
It is assumed that $q_{n}(x)$ for each $n\geq 0$ is bounded in $x$ and
vanishes outside $[0,b]$. Since $h_{n}(x;k)$ is a linear combination of $%
q_{1},\dots,q_{n-2}$ with coefficients that are polynomials in $%
u_{1},\dots,u_{n-1}$, it follows that $h_{n}(x;k)$ is bounded in $x$ and
vanishes when $x\notin [0,b]$. \qed

Recall that our aim in this paper is to solve the inverse
scattering problem for (1.1), namely to recover $Q(x,u)$ from some
scattering data involving $A(k;\varepsilon)$ and $B(k;\varepsilon)$. In
the light of (1.2) and (2.4)-(2.13), we see that our inverse
problem is equivalent to the recovery of the $q_{n-1}(x)$ for each $n\geq 1$
from some data involving $A_{n}(k) $ and $B_{n}(k) $. We will establish the
uniqueness of the recovery recursively; namely, first $q_{0}(x)$ will be
shown to be uniquely recoverable from $\{A_{1}(k),B_{1}(k)\}$, and then we
will prove the unique recovery of $q_{n-1}$ for each $n\geq 2$ from the data
set ${\Cal D}_{n}$, where we have defined
$${\Cal D}_{n}:=\{A_{n}(k),B_{n}(k),q_{0}(x),\dots,q_{n-2}(x)\},\qquad n\geq 2.
\tag{2.16}$$
As we will see, our uniqueness proof of recovery of $q_{n-1}(x)$ will rely
on the values of $A_{n}\left( k\right) $ and $B_{n}(k)$ with
large complex values of $k$. Recall that we have shown the validity of the
solution $u(x;k)$ to (1.1) with the condition in (1.6) only for real $k$,
and those of the scattering coefficients $A(k;\varepsilon)$
and $B(k;\varepsilon)$ only for real, nonzero, and bounded $k$. Hence, it is
pleasantly surprising that we can prove the uniqueness of recovery of $Q(x,u)
$ from $A(k;\varepsilon)$ and $B(k;\varepsilon)$ without needing any
extensions of the latter quantities to complex $k$ values.

\vskip 10 pt
\noindent {\bf 3. INVERSE SCATTERING TO RECOVER $q_{0}(x)$}
\vskip 3 pt

In order to solve the inverse scattering problem for (2.9), we need
some basic facts related to (2.6). In this section we list
those basic facts and refer the reader to [7-11]
for details. Associated with (2.6), let $L$ denote the unique
selfadjoint realization of $-d^{2}/dx^{2}+q_{0}(x)$
in $L_{2}({\bold R})$. Even under weaker assumptions
on $q_{0}(x)$, namely, when $q_{0}(x)$ is real valued, integrable, and
vanishing outside $[0,b]$, the following are known:

\item{(i)} $L$ has no positive or zero eigenvalues, it has no
singular-continuous spectrum, and its absolutely continuous spectrum
consists of $[0,+\infty )$. It has at most a finite number of (simple)
negative eigenvalues, and we will denote the eigenvalues by $-\kappa
_{j}^{2} $ for $j=1,...,N$.

\item{(ii)} The solution $u_{1}(x;k)$ to (2.6) satisfying
(2.7) is usually known as the Jost solution from the right and
sometimes denoted by $f_{{\text{r}}}(k,x)$. As also indicated in
Proposition~2.2, $u_{1}(x;k)$ is entire in
$k$ for each $x\in\bold R.$ From Lemma 1(ii) on page 130
of [8], it follows that for $k\in \overline{{\bold C}^{+}}$ and
$x\in [0,b]$ we have
$$|e^{ikx}u_{1}(x;k)|\leq \exp \left( b\int_{0}^{b}|q_{0}(t)|\,dt\right),
\tag{3.1}$$
and for $k\in \overline{{\bold C}^{+}}\setminus \{0\}$ and $x\in [0,b]$ we have
$$|e^{ikx}u_{1}(x;k)-1|\leq \frac{1}{|k|}\left(
\int_{0}^{b}|q_{0}(t)|\,dt\right) \exp \left(
b\int_{0}^{b}|q_{0}(z)|\,dz\right),\tag{3.2}$$
where we use ${\bold C}^{+}$ for the upper half complex plane
and put $\overline{{\bold C}^{+}}:={\bold C}^{+}\cup {\bold R}$.
Using (3.1) and (3.2), for $k\in \overline{{\bold C}^{+}}\setminus \{0\}$
and $x\in[0,b]$ we obtain
$$|e^{ik(n+1)x}u_{1}(x;k)-1|\leq \frac{(n+1)}{|k|}\left(
\int_{0}^{b}|q_{0}(t)|\,dt\right) \exp \left(
b(n+1)\int_{0}^{b}|q_{0}(z)|\,dz\right).\tag{3.3}$$

\item{(iii)} Another solution to (2.6) which we denote by
$v_{1}(x;k),$ usually known as the Jost solution from the left and sometimes
denoted by $f_{{\text{l}}}(k,x)$, satisfies the asymptotic conditions
$$v_{1}(x;k)=e^{ikx}[1+o(1)],\quad v_{1}^{\prime
}(x;k)=ik\,e^{ikx}[1+o(1)],\qquad x\to +\infty.$$
Because of the compact support property of $q_{0}(x)$ we have
$$v_{1}(x;k)=\cases
B_{1}(k)\, e^{ikx}-A_{1}(-k)\, e^{-ikx}, \qquad x\leq 0,\\
\noalign{\medskip}
e^{ikx}, \qquad x\geq b.\endcases\tag{3.4}$$
For each $x\in\bold R,$ $u_{1}(x;k)$ is entire in
$k.$ The inequalities given in (3.1)-(3.3) also hold if we replace
$e^{ikx}u_{1}(x;k)$ in them by $e^{-ikx}v_{1}(x;k)$,
which is a consequence
of Lemma 1(i) on page 130 of [8].

\item{(iv)} The scattering coefficients $A_{1}(k)$
and $B_{1}(k)$ given in (2.8) are related to the
transmission coefficient $T(k)$ and the right reflection coefficient $R(k)$
as $A_{1}(k)=R(k)/T(k)$ and $B_{1}(k)=1/T(k)$. It is
known that $kA_{1}(k)$ and $kB_{1}(k)$ are entire (cf. Proposition~2.2). The
quantity $B_{1}(k)$ is nonzero in $\overline{{\bold C}^{+}}\setminus
\{i\kappa _{j}\}_{j=1}^N$ and has simple zeros
at $k=i\kappa _{j}$ for $j=1,\dots,N$.
In the so-called generic case $B_{1}(k)$
has a simple pole at $k=0$, and in the so-called exceptional case $B_{1}(k)$
is continuous at $k=0$.

\item{(v)} The potential $q_{0}(x)$ can be recovered from
$A_{1}(k)/B_{1}(k)$ known for $k\in {\bold R}$
by any of the several methods
[7-11]. No bound state data needs to be supplied [12-17] due to the
compact support property of $q_{0}(x)$, and in fact the bound state data
(i.e. the $N$ constants $\kappa _{j}$ and the related norming constants) can
be recovered via the unique meromorphic continuation of $A_{1}(k)/B_{1}(k)$
from $k\in {\bold R}$ to $k\in {\bold C}^{+}$ and
using its poles and residues.
Actually, knowledge of $A_{1}(k)/B_{1}(k)$ in some interval on the
real $k$-axis is sufficient to recover $q_{0}(x)$
because of the uniqueness of the
meromorphic continuation.

\item{(vi)} It is known that $B_1(k)$ alone cannot uniquely
determine $q_0(x),$ and unless $B_1(k)\equiv 1$ there are
infinitely many potentials corresponding to it.
In the absence of bound states, i.e. if $N=0,$
the coefficient $A_1(k)$ uniquely determines
$q_0(x)$ provided $1/B_1(k)$ vanishes at $k=0.$
Otherwise, there are exactly two distinct potentials
corresponding to $A_1(k).$ When $N\ge 1,$ there is no
uniqueness and there are a discrete
number of potentials corresponding to $A_1(k).$ For details we refer the
reader to [18].

\item{(vii)} As long as $k^{2}$ is not in the
$L_{2}({\bold R})$-spectrum of $L$,
i.e. when $k\in {\bold C}^{+}\setminus \{i\kappa
_{j}\}_{j=1}^{N}$, the operator $(L-k^{2})^{-1}$ is a bounded
one from $L_{2}({\bold R})$
to $L_{2}({\bold R})$. The Green's function associated with (2.6),
denoted by $G(x,t;k)$, is defined as the integral kernel of
the resolvent $(L-k^{2})^{-1}$ in the sense that for any
$g\in L_{2}({\bold R})$ we have
$$\left[(L-k^{2})^{-1}g\right](x) =\int_{-\infty }^{\infty
}G(x,t;k)\,g(t)\, dt.\tag{3.5}$$
It follows that, for any $k\in {\bold C}^{+}\setminus \{i\kappa
_{j}\}_{j=1}^{N}$ [cf. (i) and (iv) above], we get
$$-G_{xx}(x,t;k)+q_{0}(x)\,G(x,t;k)=k^{2}G(x,t;k)+\delta(x-t),
\tag{3.6}$$
where $\delta(\cdot)$ is the Dirac delta distribution. With
the help of the Wronskian identity
$${2}ik{B_{1}(}k{)}=v_{1}^{\prime }(x;k)\,u_{1}(x;k)
-v_{1}(x;k)\,u_{1}^{\prime}(x;k),$$
it can be verified that
$$G(x,t;k)=\cases
-\displaystyle\frac{1}{2ik\,B_{1}(k)}\,v_{1}(t;k)\,u_{1}(x;k),
\qquad x<t,\\
\noalign{\medskip}
-\displaystyle\frac{1}{2ik\,B_{1}(k)}\,v_{1}(x;k)\,u_{1}(t;k),
\qquad  x>t.\endcases
\tag{3.7}$$
It is seen from (2.7), the second line in (3.4), and (3.7)
that, when $k\in
{\bold C}^{+}\setminus \{i\kappa _{j}\}_{j=1}^{N}$ the Green's
function $G(x,t;k)$ decays exponentially
as $x\to \pm\infty$, and it is the
only solution to (3.6) having this property.

\vskip 10 pt
\noindent {\bf 4. INVERSE SCATTERING FOR $q_{n-1}(x)$ WITH $n\geq 2$}
\vskip 3 pt

In this section we show that the recovery of $q_{n-1}(x)$ for each $n\geq 2$
is equivalent to inverting the integral equation given in (4.4) below
or the one given in (4.5). Having seen in the previous section
that $q_{0}(x)$ can be recovered uniquely from $\{A_{1}( k) ,B_{1}(k)\}$, we
proceed recursively and prove the unique recovery of $q_{n-1}(x)$
 from the data set ${\Cal D}_{n}$ defined in (2.16).

With the help of (2.6)-(2.11), we define
$$y_{n}(x;k):=u_{n}(x;k)-\frac{B_{n}(k)}
{B_{1}(k)}\,u_{1}(x;k),\qquad n\geq 2.\tag{4.1}$$
The relevant properties of $y_{n}(x;k)$ are analyzed next.

\noindent {\bf Proposition 4.1.} {\it Let $y_{n}(x;k)$ be the quantity defined in (4.1).
Then:}
\item{(i)} {\it For each $k\in {\bold C}^{+}\setminus \{i\kappa
_{j}\}_{j=1}^{N}$, where $\{i\kappa
_{j}\}_{j=1}^{N}$ is the set of zeros of $B_{1}(k)$
in ${\bold C}^{+}$, the quantity
$y_{n}(x;k)$ is a solution to (2.9), and
it is the only solution belonging to $L_{2}({\bold R})$ in $x$.
In fact, $y_{n}(\cdot;k)$ decays exponentially as $x\to\pm\infty.$}
\item{(ii)} {\it $y_{n}(x;k)$ satisfies}
$$y_{n}(x;k)=\cases -\displaystyle
\frac{B_{n}(k)}{B_{1}(k)}\,e^{-ikx},
\qquad x\leq 0, \\
\noalign{\medskip}
\left[A_{n}(k)-\displaystyle\frac{B_{n}(k)}{B_{1}(k)}\,A_{1}(k)\right]
\,e^{ikx}, \qquad x\geq b.\endcases\tag{4.2}$$
\item{(iii)} {\it In terms of the nonhomogeneous term $g_{n}(x;k)$
defined in (2.12) and the Green's function $G(x,t;k)$ given in (3.7),
we have}
$$y_{n}(x;k)=-\int_{-\infty }^{\infty }G(x,t;k)\,g_{n}(t;k)\,
dt,\qquad k\in {\bold C
}^{+}\setminus \{i\kappa _{j}\}_{j=1}^{N}.\tag{4.3}$$

\noindent PROOF: First, from Propositions~2.2 and 2.3
and the fact that the
only zeros of $B_{1}(k)$ in ${\bold C}^{+}$ occur at $k=i\kappa _{j}$
for $j=1,\dots,N,$ it
follows that $y_{n}(x;k)$ is well defined for each
$k\in {\bold C}^{+}\setminus \{i\kappa _{j}\}_{j=1}^{N}$.
Note that $y_{n}(x;k)$ solves
(2.9) because $u_{n}(x;k)$ is a solution to the same equation
and $u_{1}(x;k)$ is a solution to the corresponding homogeneous equation. We
obtain (4.2) directly from (2.7), (2.8), (2.10), and (2.11).
For each fixed $k\in {\bold C}^{+}\setminus \{i\kappa_{j}\}_{j=1}^{N}$,
since $e^{-ikx}$ and $e^{ikx}$ decay
exponentially as $x\to-\infty $ and as $x\to +\infty $,
respectively, we get the $L_{2}({\bold R})$-property
of $y_{n}(\cdot;k)$ stated in (i). Note that $y_{n}(x;k)$ is
the only $L_{2}({\bold R})$-solution to (2.9) because
the difference with any other $L_{2}({\bold R})$-solution must
satisfy (2.6); however, as indicated in (i) of Section~3, the
only $L_{2}({\bold R})$-solutions to (2.6) occur
when $k=i\kappa _{j}$ for $j=1,\dots,N$. Thus,
we have proved (i) and (ii). From the properties listed in Proposition~2.3, it
follows that $g_{n}(\cdot;k)$ is bounded and supported in $[0,b]$ and hence
belongs to $L_{2}({\bold R})$. In terms of the operator $L$ defined in
Section~3, $y_{n}(x;k)$ satisfies $(L-k^{2})y_{n}=-g_{n}$, and hence with
the help of (3.5) we obtain (4.3). \qed

In the next theorem we present the main
integral equations from which
$q_{n-1}(x)$ with $n\ge 2$ will be recovered.

\noindent {\bf Theorem 4.2.} {\it
For each $n\geq 2$, the potential $q_{n-1}(x)$ satisfies}
$$\int_{0}^{b}v_{1}(t;k)\,u_{1}(t;k)^{n}\,
q_{n-1}(t)\,dt=E_{n}(k),\qquad k\in
{\bold C},\tag{4.4}$$
$$\int_{0}^{b}u_{1}(t;k)^{n+1}q_{n-1}(t)\,dt=F_{n}(k),
\qquad k\in {\bold C},\tag{4.5}$$
{\it where $E_{n}(k)$ and $F_{n}(k)$ are completely determined by the data
${\Cal D}_{n}$ defined in (2.16), and they are given as}
$$E_{n}(k):=-2ik\,B_{n}(k)-\int_{0}^{b}v_{1}(t;k)\,h_{n}(t;k)\,dt,\tag 4.6$$
$$F_{n}(k):=2ik[B_{1}(k)\,A_{n}(k)-A_{1}(k)\,B_{n}(k)]
-\int_{0}^{b}u_{1}(t;k)\,h_{n}(t;k)\,dt.\tag{4.7}$$

\noindent PROOF: Using (2.7) and (3.4) in (3.7), we get
$$G(x,t;k)=\cases
-\displaystyle\frac{1}{2ik\,B_{1}(k)}\,e^{-ikx}v_{1}(t;k),
\qquad x\leq \min \{0,t\},
\\
\noalign{\medskip}
-\displaystyle\frac{1}{2ik\,B_{1}(k)}\,e^{ikx}u_{1}(t;k),
\qquad x\geq \max \{b,t\}.\endcases\tag{4.8}$$
Using (4.8) on the right hand side of (4.3), we obtain
$$y_{n}(x;k)=\cases
\displaystyle\frac{1}{2ikB_{1}(k)}\,\displaystyle
e^{-ikx}\int_{0}^{b}v_{1}(t;k)\,g_{n}(t;k)\,dt, \qquad
x\leq 0, \\
\noalign{\medskip}
\displaystyle\frac{1}{2ik\,B_{1}(k)}\,
e^{ikx}\int_{0}^{b}u_{1}(t;k)\,g_{n}(t;k)\,dt, \qquad
x\geq b.\endcases\tag{4.9}$$
Comparing (4.9) with (4.2), we see that
$$2ikB_{n}(k)=-\int_{0}^{b} v_1(t;k)\, g_{n}(t;k)\,dt,\qquad n\geq 2,\tag{4.10}$$
$$2ik[B_{1}(k)\,A_{n}(k)-A_{1}(k)\,B_{n}(k)]=
\int_{0}^{b}u_{1}(t;k)\,g_{n}(t;k)\,dt,\qquad n\geq 2.\tag{4.11}$$
With the help of (2.12), we rewrite
(4.10) and (4.11) as (4.4) and
(4.5), respectively. Even though we have derived
(4.4) and (4.5) only for $k\in
{\bold {C}}^{+}\setminus \{i\kappa _{j}\}_{j=1}^{N}$, with the
help analytic extensions indicated in Proposition~2.2, we see that
the former is actually valid for $k\in {\bold{C}}$ and
the latter for $k\in {\bold{C}}\setminus \{0\}$.
Furthermore, the analytic extension to $k=0$ for (4.5) is proved by showing
that the quantity $k[B_{1}(k)\,A_{n}(k)-A_{1}(k)\,B_{n}(k)]$ appearing
in (4.7) is continuous at $k=0$. In order to see this,
with the help of (2.15), as $k\to 0$ in
${\bold{C}}$ we obtain
$$2ik[B_{1}(k)\,A_{n}(k)-A_{1}(k)\,B_{n}(k)]=u_{n}^{\prime
}(x;k)[A_{1}(k)+B_{1}(k)]+O(1).$$
Although each of
$A_{1}(k)$ and $B_{1}(k)$ may have a simple pole at $k=0,$ their
sum $A_{1}(k)+B_{1}(k)$ is continuous at $k=0$ because
(2.8) implies
$$A_{1}(k)+B_{1}(k)=e^{-ikx}u_{1}(x;k)+k\,B_{1}(k)\,\frac{1-e^{-2ikx}}{k},%
\qquad x\geq b,$$
and the right hand side is $O(1)$ as $k\to 0$ in ${\bold{C}}$. \qed

With the help of Theorem~4.2, we see that our inverse scattering problem of
recovery of $q_{n-1}(x)$ for each $n\geq 2$ can be stated as follows:
Given $E_{n}(k)$ or $F_{n}(k)$, determine $q_{n-1}(x)$ for $x\in [0,b]$
by inverting either (4.4) or (4.5).

 From (2.13) and (4.6) it is seen that the data
set $\{B_{n}(k),q_{0}(x),\dots ,q_{n-1}(x)\}$
uniquely determines all the
quantities given in (4.4) except for $q_{n-1}(x)$
there, and hence one
can use (4.4) to determine $q_{n-1}(x)$
without needing $A_{n}(k)$ in the data.

\vskip 10 pt
\noindent {\bf 5. UNIQUE RECOVERY OF $q_{n-1}(x)$ WITH $n\geq 2$}
\vskip 3 pt

In this section we show that $q_{n-1}(x)$ with $n\ge 2$ can be recovered
by inverting
either (4.4) or (4.5). We will present the inversion only for (4.5) because
the inversion of (4.4) is similar.
 From (3.3) we see that we can write
$$u_{1}(t;k)^{n+1}=e^{-ik(n+1)t}\left[1+\frac{U(t;k)}{k}\right],
\qquad x\in[0,b],\quad
k\in\overline{{\bold C}^{+}}\setminus \{0\}
,\tag 5.1$$
where $U(t;k)$ is uniformly bounded
for $x\in[0,b]$ and $k\in\overline{{\bold C}^{+}};$ i.e.
$$|U(x;k)|\le M,\qquad x\in[0,b],\quad k\in\overline{{\bold C}^{+}}
,\tag 5.2$$
where $M$ is independent of $x$ and $k.$
Let us separate the real and imaginary parts of the complex $k$ variable
in ${\bold C}^{+},$ discretize the former, and keep the latter as
a fixed positive parameter by writing
$$k_{m}=\frac{2\pi m}{(n+1)b}+i\xi ,\qquad m\in {\bold Z}.\tag{5.3}$$

Let us evaluate (4.5) at $k=k_m$ and write the resulting equation
in the operator form as
$K\phi=p,$ or equivalently as
$$\int_0^b K(m,t)\,\phi(t)\,dt=p(m),\qquad m\in {\bold Z},\tag 5.4$$
where we have defined [cf. (5.1) and (5.3)]
$$K(m,t):=e^{-\xi (n+1)t}u_{1}\left( t;k\right)^{n+1}=
e^{-2\pi imt/b}\left[ 1+\frac{U(t;k_{m})}{2\pi
m(n+1)^{-1}b^{-1}+i\xi }\right],\tag 5.5$$
$$\phi(t):=e^{\xi (n+1)t}\,q_{n-1}(t),\tag 5.6$$
$$p(m):=F_n(k_m).\tag 5.7$$
Note that we suppress the dependence on $n$ and $\xi$
because the inversion of (4.5) will be done at fixed values
of $n$ and $\xi.$

When $q_0(x)\equiv 0,$ we have $u_1(t;k)\equiv e^{-ikt}$ and
$U(t;k)\equiv 0;$ in that case the operator
$K$ reduces to $K_0,$ which is given by
[cf. (5.5)]
$$(K_0g)(m)=\int_0^b K_0(m,t)\,\phi(t)\,dt
=\int_0^b e^{-2\pi imt/b}\,\phi(t)\,dt.\tag 5.8$$

The properties of $K$ and $K_0$ are analyzed next.

\noindent {\bf Theorem 5.1.} {\it
The operators $K$ and $K_0$ defined
in (5.4) and (5.8), respectively, satisfy the following:}
\item{(i)} {\it $K_0$ maps $L_2(0,b)$ into $l_2(\bold Z),$ and
its operator norm is given by
$||K_0||=\sqrt{b}.$}

\item{(ii)}  {\it $K_0^{-1}$ exists
as a map from $l_2(\bold Z)$ into $L_2(0,b),$ and
its operator norm is given by
$||K_0^{-1}||=1/\sqrt{b}.$}

\item{(iii)} {\it $K-K_0$ maps $L_2(0,b)$ into $l_2(\bold Z).$ The
operator norm of $K-K_0$ satisfies
$||K-K_0||\le \sqrt{s(\xi)},$ where $s(\xi)$
is a monotone decreasing function on
$\xi\in(0,+\infty)$ vanishing at infinity.}

\item{(iv)} {\it There exists $\xi_0>0,$ determined
by $n,$ $b,$ and $\int_0^b |q_0(t)|\,dt$ alone,
such that
the inverse operator $K^{-1}$ exists for any
$\xi>\xi_0.$}

\noindent PROOF: For the proof of (i),
note that $K_0$ is the standard Fourier series
operator and hence it maps $L_2(0,b)$ into $l_2(\bold Z).$
With the help of the completeness relation
$$\displaystyle\sum_{m\in{\bold Z}}e^{2\pi imt/b}\,e^{-2\pi imx/b}=b\cdot
\delta(t-x),\qquad t,x\in\bold R,$$
using the definition of the operator
norm, from (5.8) we obtain $||K_0||=\sqrt{b}.$
As for (ii), the inverse Fourier series
operator $K_0^{-1}$ is given by
$$\left( K_{0}^{-1}h\right)(t)=\displaystyle\frac{1}{b}\,
\displaystyle\sum_{m\in {\bold Z}}h\left( m\right) e^{2\pi imt/b},\tag 5.9$$
and its operator norm is computed in a straightforward manner with the help of
$$\int_0^b e^{2\pi imt/b}\,e^{-2\pi ijt/b}\,dt=b\cdot
\delta_{mj},\qquad m,j\in\bold Z,$$
where $\delta_{mj}$ denotes the Kronecker delta,
yielding $||K_0^{-1}||=1/\sqrt{b}.$
As for (iii), note that
the square of the operator norm of $K-K_0$ is defined as
$$||K-K_{0}||^{2}:=\sup_{||g||=1}\left\| \left(
K-K_{0}\right) g\right\| ^{2}=\sup_{\left\| g\right\| =1}\sum_{m\in {\bold Z}%
}\left| \left[ \left( K-K_{0}\right) g\right] \left( m\right) \right| ^{2},
$$
where $||g||$ denotes the norm of $g$ in $L_{2}(0,b)$. With
the help of (5.3)-(5.5) and (5.8) we get
$$\left\| K-K_{0}\right\|^{2}=\sup_{\left\| g\right\| ^{2}=1}\sum_{m\in {\bold
Z}}\left| \int_{0}^{b}\frac{U(t;k_{m})}{2\pi
m(n+1)^{-1}b^{-1}+i\xi}\,\phi(t)\,dt\right|^{2}.\tag 5.10$$
Using the Cauchy-Schwarz inequality on (5.10), $||g||^{2}=1,$
and (5.2), we obtain
$$\aligned
||K-K_{0}||^{2}&\leq \displaystyle\sum_{m\in {\bold Z}}\int_{0}^{b}\left|
\frac{U(t;k_m)}{2\pi m(n+1)^{-1}b^{-1}+i\xi }\right|^{2}\,dt \\
\noalign{\medskip}
&\leq M^{2}b\displaystyle\sum_{m\in {\bold Z}}\frac{1}{4\pi
^{2}m^{2}(n+1)^{-2}b^{-2}+\xi^{2}},\endaligned$$
and since the summation in the last term can be evaluated
in a closed form, we obtain
$$||K-K_{0}||^{2}\leq \displaystyle\frac{M^{2}b^{2}(n+1)}{2\xi}\,
\coth \left( \frac{(n+1)b\xi}{2}\right),
\qquad \xi>0,\quad n\ge 2.\tag 5.11$$
Note that $\coth(x)$ is monotone decreasing
on $x\in(0,+\infty).$ Using (3.3), (5.2), and (5.11), we see that
$||K-K_{0}||\le \sqrt{s(\xi)},$ where
$$s(\xi):=\displaystyle\frac{b^2\,(n+1)^3}{2\xi}\left(
\int_0^b |q_0(x)|\,dx\right)^2\,
\coth \left( \frac{(n+1)b\xi}{2}\right)
\,\exp\left(2b(n+1) \int_0^b |q_0(x)|\,dx\right).$$
Next, we need to prove (iv).
Writing
$$K=K_0+(K-K_0)=K_0[I+K_0^{-1}(K-K_0)],$$
we see that $K^{-1}$ can be evaluated as a Neumann series
$$K^{-1}=[I+K_0^{-1}(K-K_0)]^{-1}K_0^{-1}=
\displaystyle\sum_{j=0}^\infty (-1)^j
[K_0^{-1}(K-K_0)]^jK_0^{-1},$$
whose convergence is assured by choosing
$||K_0^{-1}||\,||(K-K_0)||<1.$ From (ii) and (iii), we see
that this is achieved by choosing $\xi_0$ as the unique solution
to $s(\xi_0)=b.$ Then, for any $\xi>\xi_0,$
the existence of $K^{-1}$ is assured. \qed

Having shown that $K$ is invertible for sufficiently
large $\xi$ values, we can recover $q_{n-1}(x)$ with the help of
(5.4), (5.6), and (5.7) as
$$q_{n-1}(x)=e^{-\xi(n+1)x}\,(K^{-1}F_n)(x),\qquad
x\in[0,b],\quad n\ge 2.\tag 5.12$$

Let us now summarize the reconstruction
in the solution of the inverse scattering problem for (1.1).
Assume that we are given the scattering data
consisting of $A(k;\varepsilon)$ and $B(k;\varepsilon)$ for
all $k$ in some subinterval
of the positive $k$-axis and for all
$|\varepsilon|\le \delta,$ where $\delta$
is some positive number (no matter how small).
Our aim is to recover the potential $Q(x,u)$ corresponding to
the given scattering data.
\item{(i)} Using (2.5), obtain $A_n(k)$ and $B_n(k)$ for $n\ge 1.$
Note that $kA_n(k)$ and $kB_n(k)$ have entire extensions to $\bold C.$
\item{(ii)} Recover $q_0(x)$ and $u_1(x;k)$ from
$\{A_1(k),B_1(k)\}$ by using any one of
the available inversion methods described in (v) of Section~3.
\item{(iii)} Recover $q_{n-1}(x)$ and $u_n(x;k)$
for $n\ge 2$ by using the data $\Cal D_n$ given in (2.16) in
a recursive way. The recovery of $q_{n-1}(x)$
is achieved by inverting either (4.4) or (4.5). The recovery
of $u_n(x;k)$ amounts to solving the linear equation (2.9) with
the condition in (2.10).
\item{(iv)} Having recovered all the $q_{n-1}(x)$ for $n\ge 1,$ we obtain
$Q(x,u)$ via (1.2).

The uniqueness of the recovery of $Q(x,u)$ from
the data $\{A(k;\varepsilon),B(k;\varepsilon)\}$ is summarized next.

\noindent {\bf Theorem 5.2.} {\it Consider the data
consisting of $A(k;\varepsilon)$ and $B(k;\varepsilon)$
given for
all $k$ in some subinterval
of the positive $k$-axis and
for all $|\varepsilon|\le \delta,$ where $\delta$ is
some positive number (no matter how small). Further
assume that there exists a potential $Q(x,u)$
corresponding to this data. Then $Q(x,u)$ is the only potential
corresponding to that data.}

\vskip 10 pt
\noindent {\bf 6. EXAMPLES}
\vskip 3 pt

In this section we illustrate the direct and inverse problems
for (1.1) in a special case with some explicit examples.
Assume that $Q(x,u)$ given in (1.2) has the form $Q(x,u)=q_2(x)\,u^2,$
i.e. $q_j(x)\equiv 0$ for $j\ge 0$ except when $j=2.$
 From (2.6)-(2.8) we get $u_1(x;k)=e^{-ikx}$ for $x\in\bold R,$
$A_1(k)=0,$ $B_1(k)=1.$
 From (2.9)-(2.13) with $n=2,$ we obtain $u_2(x;k)=0$ for $x\in\bold R,$
$A_2(k)=B_2(k)=0.$ Using (2.9)-(2.11) with $n=3,$ we see that
$u_3(x;k)$ satisfies
$$u_3''(x;k)+k^2u_3(x;k)=q_2(x)\,e^{-3ikx},\qquad x\in\bold R$$
with the initial conditions $u_3(0;k)=u_3'(0;k)=0.$ Using
variation of parameters, we get
$$u_3(x;k)=\displaystyle\frac{1}{2ik}\,e^{ikx}\displaystyle\int_0^x q_2(t)\,
e^{-4ikt}dt-\displaystyle\frac{1}{2ik}\,e^{-ikx}\displaystyle\int_0^x q_2(t)\,
e^{-2ikt}dt,\qquad x\in\bold R.\tag 6.1$$ Comparing (6.1) with
(2.11) we see that
$$A_3(k)=\displaystyle\frac{1}{2ik}\displaystyle\int_0^b
q_2(t)\,e^{-4ikt}dt,\quad B_3(k)=-\displaystyle\frac{1}{2ik}\displaystyle\int_0^b
q_2(t)\,e^{-2ikt}dt.\tag 6.2$$
Thus, we see that $A_3(k)$ and
$B_3(k)$ are related to each other as $A_3(k)=-2\,B_3(2k).$ With
the help of $\int_{-\infty}^\infty
e^{ik\alpha}dk=2\pi\delta(\alpha),$ the potential $q_2(x)$ is
uniquely recovered from $A_3(k)$ or $B_3(k)$ given in (6.2) as
$$q_2(x)=\displaystyle\frac{4i}{\pi}\displaystyle\int_{-\infty}^\infty
k\,A_3(k)\,e^{4ikx}dk=-\displaystyle\frac{2i}{\pi}\displaystyle\int_{-\infty}^\infty
k\,B_3(k)\,e^{2ikx}dk.\tag 6.3$$

Alternatively, $q_2(x)$ can be recovered by using the method of Section~5
as follows. Let us suppose that we are given $A_3(k).$
Using (2.14) and (4.7) we see that
$F_3(k)=2ikA_3(k).$ Thus, with the help of (5.3) and (5.7) we get
$$p(m)=(im\pi/b-2\xi)\,A_3(i\xi+m\pi/2b),\qquad m\in\bold Z.$$
Since $q_0(x)\equiv 0,$ we have $K=K_0,$ where $K$ and $K_0$ are
the operators appearing in (5.4) and (5.8), respectively.
Thus, for any $\xi>0,$ with the help of
(5.9) and (5.12), we construct $q_2(x)$ explicitly on $[0,b]$ via
$$q_2(x)=\displaystyle\frac{1}{b^2}\,e^{-\xi(n+1)x}\,
\displaystyle\sum_{m\in {\bold Z}} e^{2\pi imx/b}\,(im\pi-2b\xi)\,A_3(i\xi+m\pi/2b).
\tag 6.4$$

In particular, if $q_2(x)$ is a constant on the
interval $[0,b],$ say $q_2(x)=\gamma,$ from (6.2) we get
$$A_3(k)=-\displaystyle\frac{\gamma}{8k^2}\left(1-e^{-4ikb}\right),
\quad
B_3(k)=\displaystyle\frac{\gamma}{4k^2}\left(1-e^{-2ikb}\right).\tag 6.5$$
Conversely, if $A_3(k)$ and $B_3(k)$ are as in (6.5), then
with the help of
$$\int_{-\infty}^\infty \displaystyle\frac{e^{ika}}{k}\,dk=i\pi\,\text{sgn}(a),$$
where $\text{sgn}(a)$ denotes
the signature function, we recover $q_2(x)=\gamma$ via (6.3).
Alternatively, $q_2(x)$ can be recovered by using (6.4).

In another particular case, in which $q_2(x)=e^{\alpha x}$ on
the interval $[0,b],$ where $\alpha$ is a constant, from (6.2) we get
$$A_3(k)=\displaystyle\frac{e^{(\alpha-4ik)b}-1}{2ik(\alpha-4ik)},
\quad
B_3(k)=-\displaystyle\frac{e^{(\alpha-2ik)b}-1}{2ik(\alpha-2ik)}.\tag 6.6$$
Conversely, if $A_3(k)$ and $B_3(k)$ are as in (6.6), then
we recover $q_2(x)=e^{\alpha x}$ via (6.3) with the help of
a contour integration. Alternatively, we
can recover $q_2(x)$ by using (6.4).

\vskip 10 pt

\noindent {\bf Acknowledgment.} The research leading to this
article was supported in part by the National Science Foundation
under grant DMS-0204437 and the Department of Energy under grant
DE-FG02-01ER45951.

\vskip 10 pt

\noindent {\bf{References}}

\item{[1]} R. Weder,
{\it Inverse scattering for the non-linear Schr\"{o}dinger equation:
Reconstruction of the potential and the non-linearity,}
Math. Meth. Appl. Sci. {\bf 24}, 245--254 (2001).

\item{[2]} R. Weder,
{\it Inverse scattering on the line for the nonlinear Klein-Gordon
equation with a potential,}
J. Math. Anal. Appl. {\bf 252}, 102--123 (2000).

\item{[3]} C. S. Morawetz and W. A. Strauss, {\it
On a nonlinear scattering operator,}
Commun. Pure Appl. Math. {\bf 26}, 47--54 (1973).

\item{[4]} A. Bachelot,
{\it Inverse scattering problem for the nonlinear Klein-Gordon equation,}
In: C. Bardos et al. (eds.),
{\it Contributions to nonlinear partial differential equations (Madrid, 1981)},
Res. Notes in Math. {\bf 89},
Pitman, Boston, MA, 1983, pp. 7--15.

\item{[5]} R. Weder,
{\it Inverse scattering for the nonlinear Schr\"{o}dinger
equation. II. Reconstruction of the potential and the
nonlinearity in the multidimensional case,}
Proc. Amer. Math. Soc. {\bf 129}, 3637--3645 (2001).

\item{[6]} E. A. Coddington and N. Levinson, {\it Theory of ordinary differential equations,}
Robert E. Krieger Publ. Co., Malabar, Florida, 1987.

\item{[7]} L. D. Faddeev, {\it
Properties of the $S$-matrix of the one-dimensional Schr\"{o}dinger
equation,} Amer. Math. Soc. Transl. (ser. 2) {\bf 65}, 139--166 (1967).

\item{[8]} P. Deift and E. Trubowitz, {\it
Inverse scattering on the line,} Commun. Pure Appl. Math. {\bf 32},
121--251 (1979).

\item{[9]}  V.\; A.\; Marchenko, {\it Sturm-Liouville operators and
applications,} Birk\-h\"au\-ser, Basel, 1986.

\item{[10]} K. Chadan and P. C. Sabatier, {\it Inverse problems in quantum
scattering theory,} 2nd ed., Springer, New York, 1989.

\item{[11]} T. Aktosun and M. Klaus, {\it
Chapter 2.2.4, Inverse theory: problem on the line,} In: E. R.
Pike and P. C. Sabatier (eds.), {\it Scattering,} Academic Press,
London, 2001, pp. 770--785.

\item{[12]} N. N. Novikova and V. M. Markushevich,
{\it Uniqueness of the solution of the one-dimensional problem of
scattering for potentials located on the positive semiaxis,}
Comput. Seismology {\bf 18}, 164--172 (1987).

\item{[13]} T. Aktosun, M. Klaus, and C. van der Mee C,
{\it On the Riemann-Hilbert problem for the one-dimensional
Schr\"odinger equation,} J. Math. Phys. {\bf 34}, 2651--2690
(1993).

\item{[14]} T. Aktosun,
{\it Bound states and inverse scattering for the Schr\"odinger
equation in one dimension,} J. Math. Phys. {\bf 35}, 6231--6236
(1994).

\item{[15]} B. Gr\'ebert and R. Weder,
{\it Reconstruction of a potential on the line that is a priori
known on the half line,} SIAM J. Appl. Math. {\bf 55}, 242--254
(1995).

\item{[16]} T. Aktosun,
{\it Inverse Schr\"odinger scattering on the line with partial
knowledge of the potential,} SIAM J. Appl. Math. {\bf 56},
219--231 (1996).

\item{[17]} F. Gesztesy and B. Simon,
{\it Inverse spectral analysis with partial information on the
potential. I. The case of an a.c. component in the spectrum,}
Helv. Phys. Acta {\bf 70}, 66--71 (1997).

\item{[18]} T. Aktosun and V. G. Papanicolaou, {\it
Recovery of a potential from the ratio of reflection and transmission coefficients,} J. Math.
Phys. {\bf 44}, 4875--4883 (2003).

\end